\documentclass[a4paper,11pt]{amsart}
\usepackage{graphicx}
\begin{document}
\hyphenation{gra-vi-ta-tio-nal re-la-ti-vi-ty Gaus-sian
re-fe-ren-ce re-la-ti-ve gra-vi-ta-tion Schwarz-schild
ac-cor-dingly gra-vi-ta-tio-nal-ly re-la-ti-vi-stic pro-du-cing
de-ri-va-ti-ve ge-ne-ral ex-pli-citly des-cri-bed ma-the-ma-ti-cal
de-si-gnan-do-si coe-ren-za pro-blem gra-vi-ta-ting geo-de-sic
per-ga-mon cos-mo-lo-gi-cal gra-vity cor-res-pon-ding
de-fi-ni-tion phy-si-ka-li-schen ma-the-ma-ti-sches ge-ra-de
Sze-keres}
\title[On repulsive gravitational actions]
{{\bf On repulsive gravitational actions}}

\author[Angelo Loinger]{Angelo Loinger}
\address{A.L. -- Dipartimento di Fisica, Universit\`a di Milano, Via
Celoria, 16 - 20133 Milano (Italy)}
\author[Tiziana Marsico]{Tiziana Marsico}
\address{T.M. -- Liceo Classico ``G. Berchet'', Via della Commenda, 26 - 20122 Milano (Italy)}
\email{angelo.loinger@mi.infn.it} \email{martiz64@libero.it}

\vskip0.50cm

\begin{abstract}
In particular instances and in particular regions the Einsteinian
gravity exerts a \textbf{\emph{repulsive}} action -- and
\emph{without} any cosmological term. This fact could have an
interest for the explanation of the dark energy, and for the
gravitational collapses.
\end{abstract}

\maketitle


General relativity.

\normalsize

\vskip1.20cm \noindent \textbf{1}. -- In a recent paper (see
sects. \textbf{5} and \textbf{5bis} of \cite{1}) we have
emphasized, reconsidering some lucid and straightforward
computations by Hilbert \cite{2}, that in particular instances and
in particular regions the Einsteinian gravity -- \emph{without}
any cosmological term -- becomes of a \textbf{\emph{repulsive}}
kind. We give now an impressive illustration of this fact by
exhibiting six diagrams (five concerning test-particles and one
concerning light-rays, moving in a Schwarzschild field) which are
specially enlightening.

\vskip0.80cm \noindent \textbf{2}. -- Of course, Hilbert \cite{2}
uses the \emph{standard} form of solution to Schwarzschild
problem, which was discovered by him, by Droste, and by Weyl,
quite independently: $g_{00}=1-2m/r$; $g_{00}^{-1}=g_{rr}$;
$g_{\vartheta\vartheta}=r^{2}$;
$g_{\varphi\varphi}=r^{2}\sin^{2}\vartheta$, where $m\equiv
GM/c^{2}$ and $M$ is the mass of the gravitating centre;
$g_{jk}=0$ for $j\neq k$. Remark that for $r>2m$ the standard form
is \emph{diffeomorphic} to the \textbf{\emph{original}}
Schwarzschild's form and to Brillouin's form, that hold for $r>0$
(see, \emph{e.g.}, \cite{1}, and the Appendix).

\par Hilbert gives a full treatment of the geodesic lines of
test-particles and light-rays in a Schwarzschild field. We
summarize here (see also \cite{1}) the Hilbertian results
regarding the circular paths and the radial ones. For a simple
reason of \emph{physical reality} -- and in accordance with ideas
of Einstein and Hilbert -- we assume that the gravitating centre
is an \emph{extended} spherical body with the minimal radius
$(9/8) \, 2m$ \cite{3}.

\par The \emph{circular} geodesics of the test-particles are restricted
by the following inequalities (Hilbert puts $c=1$, and his
$\alpha$ coincides with $2m$):

\begin{equation} \label{eq:one}
r > \frac{3}{2} (2m) \quad \left[>\frac{9}{8} (2m)\right] \quad,
\end{equation}

\begin{equation} \label{eq:two}
\frac{v}{c} < \frac{1}{\sqrt3} \quad,
\end{equation}

where $v=c \, (m/r)^{1/2}$ is the linear velocity. These relations
tell us that here the Einsteinian gravity acts
\textbf{\emph{repulsively}} for \emph{small} values of $r$.

\par For the \emph{circular} trajectories of the light-rays the
coordinate-radius $r$ is equal to $(3/2)2m$, with a velocity
$v=c/\sqrt3$.

\par In 1939 Einstein \cite{4}, employing a system of isotropic
coordinates, rediscovered relation (\ref{eq:one}); he found that
(his $\mu$ is equal to our $m$):

\begin{equation} \label{eq:three}
r> \frac{m}{2} \left(2+\sqrt3\right) \quad,
\end{equation}

where $r$ is now the radial isotropic coordinate. The concordance
of (\ref{eq:one}) with (\ref{eq:three}) is a straightforward
consequence of the passage from standard $r$ to isotropic $r$:

\begin{equation} \label{eq:four}
r \rightarrow \left( 1+ \frac{m}{2r}\right)^{2}r \quad. \quad-
\end{equation}

In the standard coordinates the differential equation of the
\emph{radial} motions is:

\begin{equation} \label{eq:five}
\frac{1}{c^{2}} \frac{\textrm{d}^{2}r}{\textrm{d}t^{2}}
-\frac{3}{2} \,\frac{2m}{r(r-2m)}
\left(\frac{\textrm{d}r}{c\textrm{d}t}\right)^{2} +
\frac{m(r-2m)}{r^{3}} = 0\quad;
\end{equation}

with the integral:

\begin{equation} \label{eq:six}
\left(\frac{\textrm{d}r}{c \, \textrm{d}t}\right)^{2} =
\left(\frac{r-2m}{r}\right)^{2} + A
\left(\frac{r-2m}{r}\right)^{3} \quad;
\end{equation}

the constant $A$ is zero for the light-rays, negative for the
material particles, and such that $(2/3) \leq |A| \leq 1$.

\par Eqs. (\ref{eq:five}) and (\ref{eq:six}) tell us that the
acceleration is negative (attractive gravity) or positive
(\textbf{\emph{repulsive}} gravity) where, respectively:

\begin{equation} \label{eq:seven}
\left|\frac{\textrm{d}r}{c \, \textrm{d}t}\right| <
\frac{1}{\sqrt3} \,  \frac{r-2m}{r} \quad,
\end{equation}

\begin{equation} \label{eq:eight}
\left|\frac{\textrm{d}r}{c \, \textrm{d}t}\right| >
\frac{1}{\sqrt3}  \, \frac{r-2m}{r} \quad.
\end{equation}

For the \emph{radial} motions of the light-rays we have from eq.
(\ref{eq:six}) with $A=0$:

\begin{equation} \label{eq:nine}
\left|\frac{\textrm{d}r}{c \, \textrm{d}t}\right| = \frac{r-2m}{r}
\quad,
\end{equation}

and therefore the light is everywhere \emph{repulsed} by the
gravitating body. (see  inequality (\ref{eq:eight})). If, in
particular, a light-ray starts from $r=\infty$ with a velocity
$c$, it arrives at $r=(9/8)2m$ with velocity $(1/9)c$.

\par Putting $x:= r/(2m)$ and $y:=
(\textrm{d}r/c \, \textrm{d}t)^{2}$, eq. (\ref{eq:six}) can be
rewritten as follows:

\begin{equation} \label{eq:ten}
y(x) = \left(\frac{x-1}{x}\right)^{2} \left(1- |A|
\frac{x-1}{x}\right) \quad, \quad (1<x<\infty) \quad.
\end{equation}

\par Figs. 1$\div$5 represent five diagrams of eq. (\ref{eq:ten})
for the following values of $|A|$: $1$; $0.9$; $0.8$; $0.7$;
$2/3$. Fig. \ref{eq:six} gives the diagram of eq. (\ref{eq:ten})
for $A=0$. In these diagrams the variable $x$ goes from $(9/8)$ to
infinite. The regions in which the gravitation acts repulsively
are quite evident. Remark that in the instances of Fig.
\ref{eq:five} (test-particles) and Fig. \ref{eq:six} (light-rays)
the gravitation is \emph{everywhere} repulsive.

\newpage
\begin{figure}[!hbp]
\begin{center}
\includegraphics[width=1.0\textwidth]{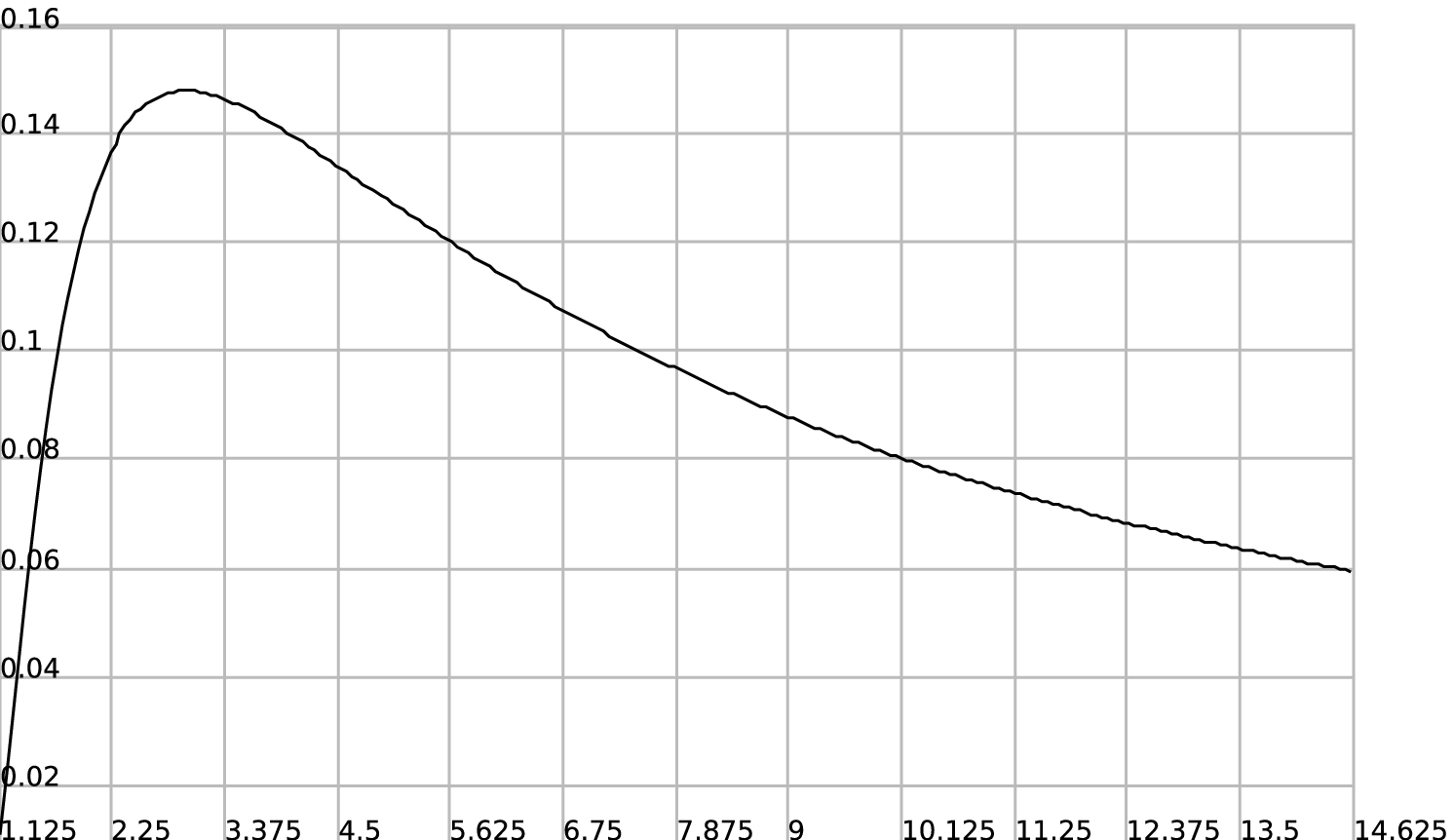}
\caption{Diagram of $y(x)=[(x-1)/x]^{2}[1-(x-1)/x]$ for some
values of $x$; $(9/8)\leq x <+\infty$; $\max(3.0;4/27)$;
$[y(9/8)]^{1/2}=2\sqrt2 /27$.} 
\vskip1.00cm
\includegraphics[width=1.0\textwidth]{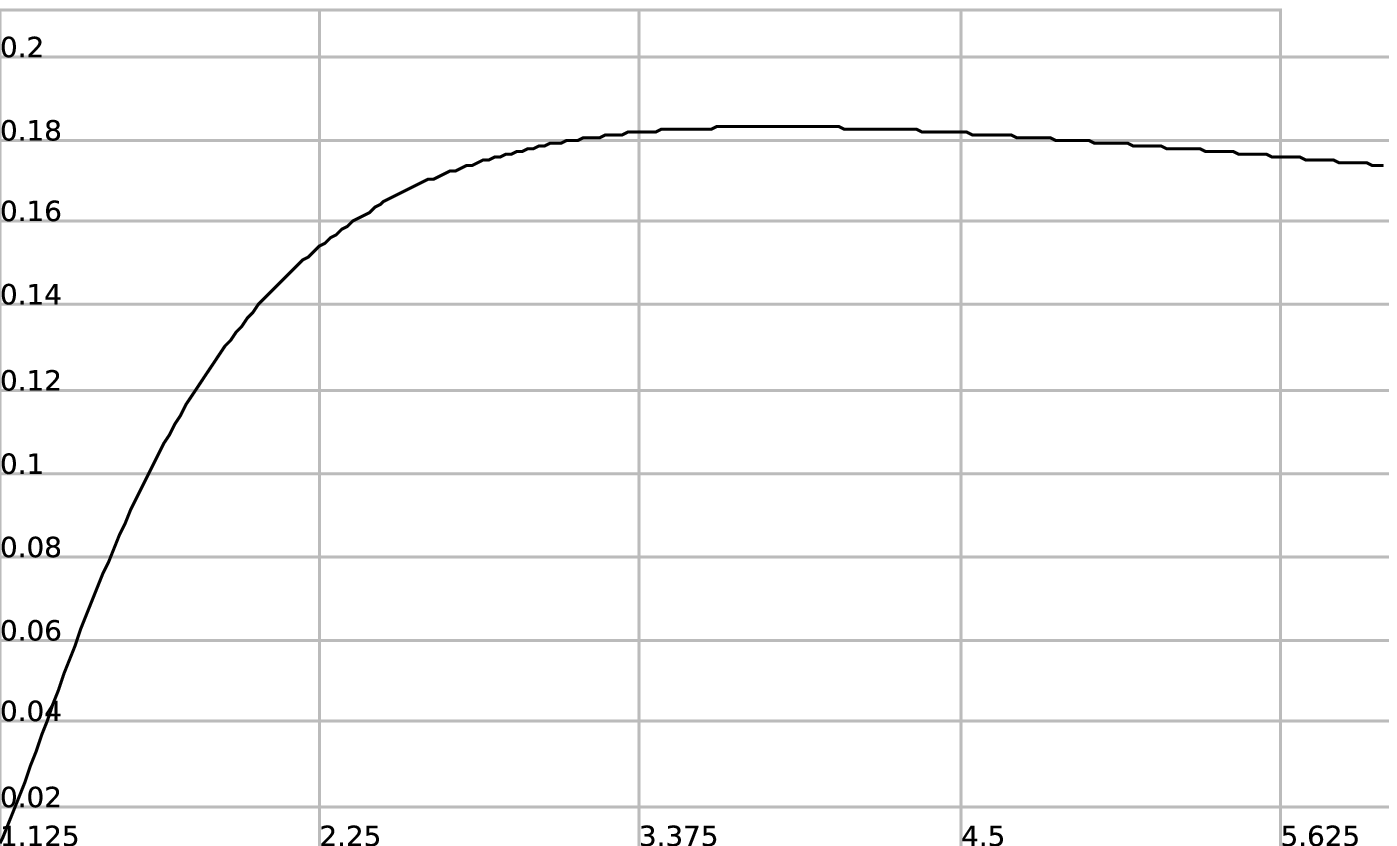}
\caption{Diagram of $y(x)=[(x-1)/x]^{2}[1-0.9*(x-1)/x]$ for some
values of $x$; $(9/8)\leq x <+\infty$; $\max(3.75;0.182844)$;
$[y(9/8)]^{1/2}=0.105409$.}
\end{center}
\end{figure}

\newpage
\begin{figure}[!hbp]
\begin{center}
\includegraphics[width=1.0\textwidth]{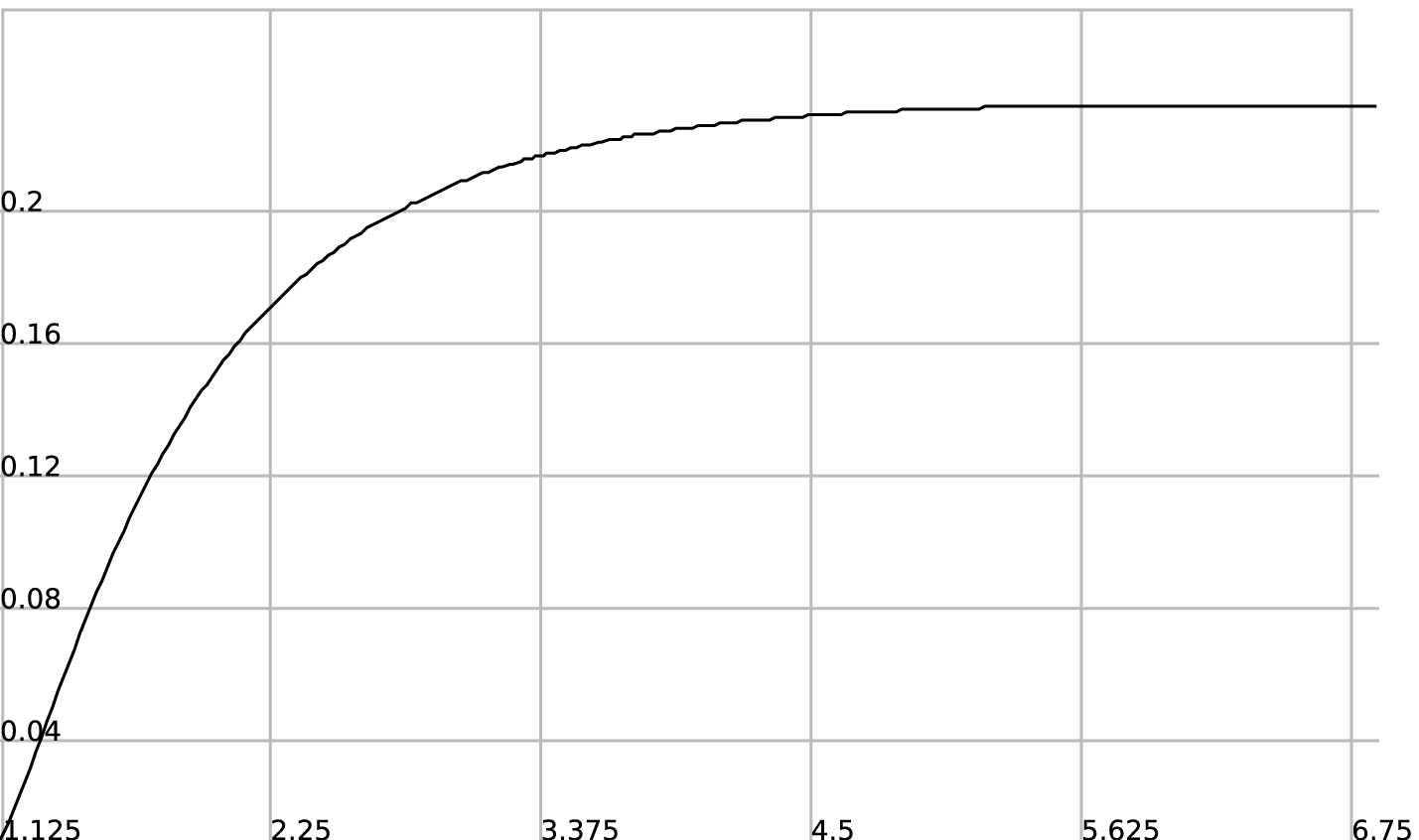}
\caption{Diagram of $y(x)=[(x-1)/x]^{2}[1-0.8*(x-1)/x]$ for some
values of $x$; $(9/8)\leq x <+\infty$; $\max(6.0;0.231481)$;
$[y(9/8)]^{1/2}=0.106058$.} \vskip1.00cm
\includegraphics[width=1.0\textwidth]{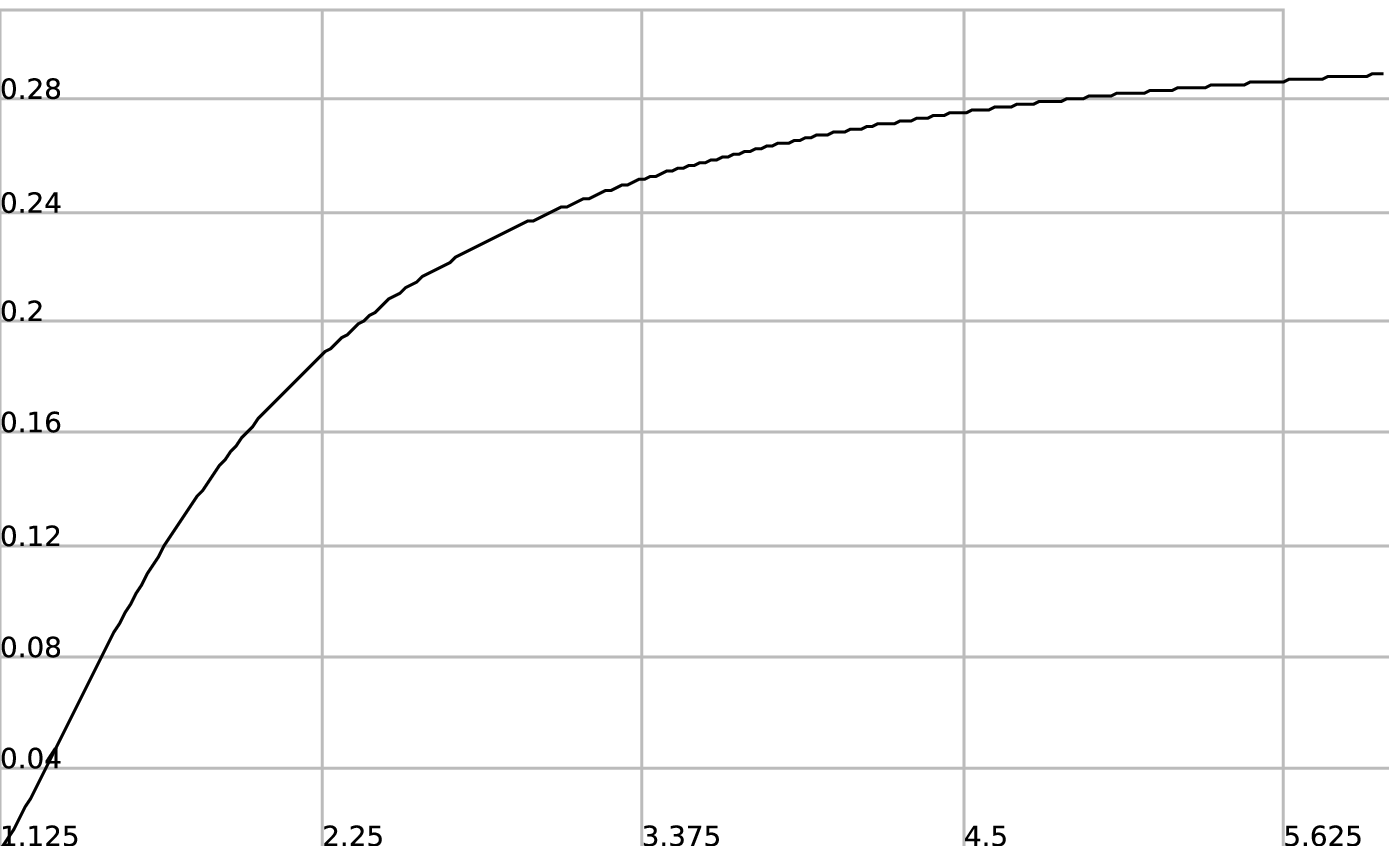}
\caption{Diagram of $y(x)=[(x-1)/x]^{2}[1-0.7*(x-1)/x]$ for some
values of $x$; $(9/8)\leq x <+\infty$; $\max(21.0;0.302343)$;
$[y(9/8)]^{1/2}=0.106703$.}
\end{center}
\end{figure}

\newpage
\begin{figure}[!hbp]
\begin{center}
\includegraphics[width=1.0\textwidth]{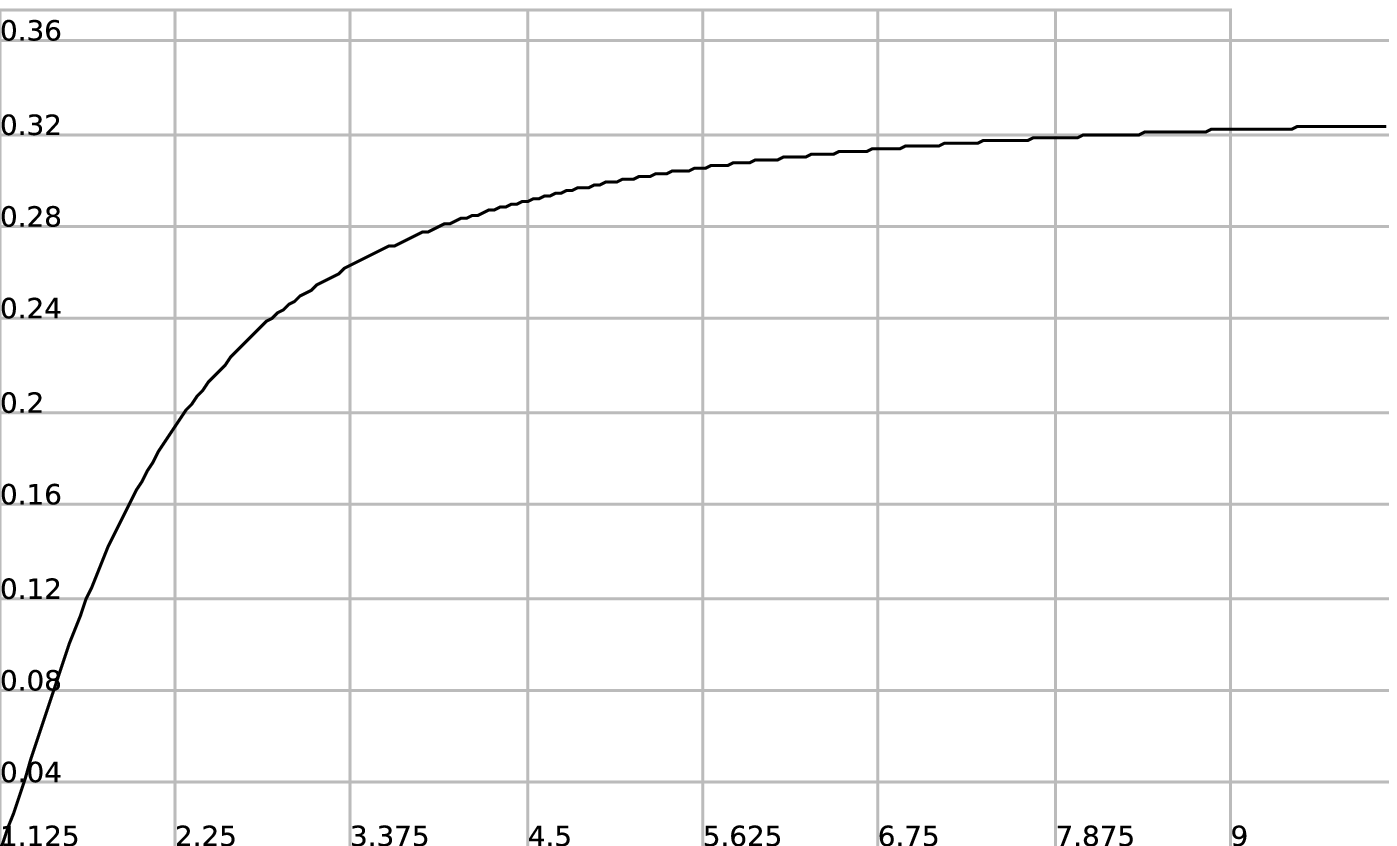}
\caption{Diagram of $y(x)=[(x-1)/x]^{2}[1-(2/3)*(x-1)/x]$ for some
values of $x$; $(9/8)\leq x <+\infty$; $\max(+\infty,1/3)$;
$[y(9/8)]^{1/2}=0.106917$.}
 \vskip1.00cm
\includegraphics[width=1.0\textwidth]{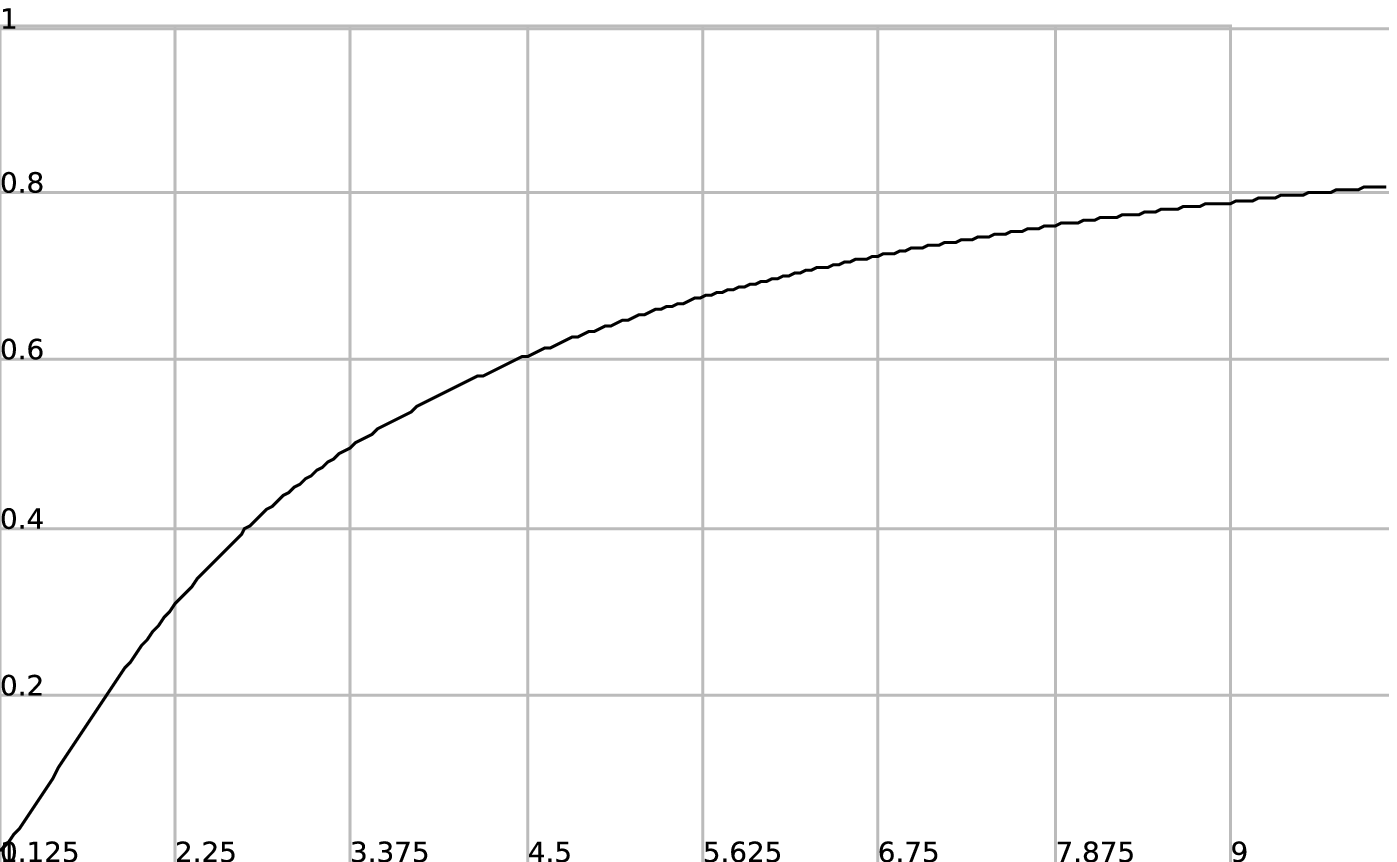}
\caption[5bis]{Diagram of $y(x)=[(x-1)/x]^{2}$ for some values of
$x$; $(9/8)\leq x <+\infty$; $\max(+\infty;1.0)$;
$[y(9/8)]^{1/2}=1 /9$.}
\end{center}
\end{figure}

\newpage
 \vskip0.80cm \noindent \textbf{3}. -- It is clear that a
 cosmological model incorporating the existence of the above
 \emph{repulsive} actions could explain the so-called \textbf{\emph{dark
 energy}}, \emph{i.e.}  the energy that is responsible for the
 \emph{accelerated} expansion of the universe \cite{5}.

 \par Further, the repulsive gravity could play a role in the
 dynamics of \emph{gravitational collapses}.

 \vskip0.80cm \noindent \textbf{4}. -- Till now people have tried
 to explain the dark energy by means of Friedmann model with a
 cosmological term, or with a ``quintessence'' \cite{5}.

 \par However, Friedmann model is rather poor, and is
 \emph{isomorphic} to a corresponding \textbf{\emph{Newtonian}}
 model \cite{6}. Accordingly, it cannot give a physically sensible
 explanation of the dark energy.

\vskip0.80cm
\begin{center}
\noindent \small \emph{\textbf{APPENDIX}}
\end{center}

\normalsize \noindent
\par The Hilbertian conception about the singularities of the
metric tensor of GR is illustrated by these considerations
concerning the Schwarzschild problem \cite{2}:

\par ``F\"{u}r $\alpha$ $[\equiv 2m] \neq 0$ erweisen sich $r=0$ und
bei positivem $\alpha$ auch $r=\alpha$ als solche Stellen, an
denen die Ma\ss{}bestimmung nicht regul\"{a}r ist. Dabei nenne ich
eine Ma\ss{}bestimmung oder oder ein Gravitationsfeld $g_{\mu\nu}$
an einer Stelle \emph{regul\"{a}r}, wenn es m\"{o}glich ist, durch
umkehrbar eindeutige Transformation ein solches Koordinatensystem
einzuf\"{u}hren, da\ss{}  f\"{u}r dieses die entsprechenden
Funktionen $g'_{\mu\nu}$ an jener Stelle regul\"{a}r, d.h. in ihr
und in ihrer Umgebung stetig und beliebig oft differenzierbar sind
und eine von Null verschiedene Determinante $g'$ haben.

\par Obwohl nach meiner Auffassung nur regul\"{a}re Losungen der
physikalischen Grundgleichungen die Wirklichkeit unmittelbar
darstellen, so sind doch gerade die L\"{o}sungen mit nicht
regul\"{a}ren Stellen ein wichtiges mathematisches Mittel zur
Ann\"{a}herung $[$approximation$]$ an charakteristische
regul\"{a}re L\"{o}sungen  -- und in diesem Sinne ist nach dem
Vorgange von \textsc{Einstein} $[$\emph{Berl. Ber.}, (1915) 831$]$
und \textsc{Schwarzschild} $[$\emph{Berl. Ber.}, (1916) 189; an
English version in \emph{arXiv:\-physics/9905030} (May 12th,
1999), and in \emph{Gen. Rel. Grav.}, \textbf{35} (2003) 951$]$
die f\"{u}r $r=0$ und $r=\alpha$ nicht regul\"{a}re
Ma\ss{}be\-stimmung (35) $[: G(\textrm{d}r, \textrm{d}\vartheta,
\textrm{d}\varphi, \textrm{d}t)=
\{r/(r-\alpha)\}\textrm{d}r^{2}+r^{2}\textrm{d}\vartheta^{2}+r^{2}\sin^{2}\vartheta
\textrm{d}\varphi^{2} - \{(r-\alpha)/r\}\textrm{d}t^{2}]$ als
Ausdruck der Gravitation einer in der Umgebung des Nullpunktes
zentrisch-symmetrisch verteilten $[$and therefore
\emph{extended}$]$ Masse anzusehen$^{1}$.'' The footnote$^{1}$
says: ``Die Stellen $r=\alpha$ nach dem Nullpunkt zu
transformieren, wie es Schwarzschild $[$\emph{loc. cit.}$]$ tut,
ist meiner Meinung nach nicht zu empfehlen $[$not to be
recommended$]$; die Schwarzschildsche Transformation ist
\"{u}berdies nicht die einfachste, die diesen Zweck erreicht.''

\par \textbf{\emph{A comment}}. - \emph{i}) The above definition
of regularity of the metric tensor is chiefly significant so far
as \emph{physical reality} is concerned. In particular, it tells
us that the well-known coordinate system by Kruskal and Szekeres
is destitute of any physical value:  indeed, the derivatives
$\partial u/\partial r$ and $\partial v/\partial r$ of Kruskal's
coordinates $u$ and $v$ are singular at $r=2m$: this means that
the singularity at $r=2m$ of the standard interval has been
``incorporated'' in the new coordinates. Consequently, the
Kruskal-Szekeres form of solution does \emph{not} give a
\emph{proper} transformation of standard form. --
\newpage \emph{ii}) Hilbert thinks that the singularities at $r=0$
and at $r=2m$ are only an \emph{approximate description} of a
small \textbf{\emph{extended}} gravitating mass having its centre
at $r=0$; the first statement of footnote$^{1}$ reveals its true
meaning if we take into account this consideration; it does not
represent a disowning of Schwarzschild's procedure. On the other
hand, the above quoted Schwarzschild's memoir concerns ``das
Gravitationsfeld eines Massenpunkt\-es'', whereas Hilbert's
treatment intends to find the Einsteinian field outside a
\emph{generic} spherosymmetrical distribution of matter.

\par The second statement of footnote$^{1}$, according to which
there exist choices of the coordinate system characterized by a
unique singularity of metric tensor at $r=0$, and that are
\emph{simpler} than Schwarzschild's choice, finds a well-known
example in Brillouin's choice $[$\emph{Journ. Phys. Radium},
\textbf{23} (1923) 43; an English version in
\emph{arXiv:physics/0002009} (February 3rd, 2000)$]$, which can be
obtained from standard $r$ with the substitution $r \rightarrow
r+2m$. (The corresponding substitution for Schwarzschild's choice
is: $r \rightarrow [r^{3}+(2m)^{3}]^{1/3}$. Brillouin's and
Schwarzschild's forms of solution are \emph{maximally extended}).

\par A last remark. After 1960 innumerable papers on the
singularities of the metric tensor have been written, and various
definitions of regularity of the solutions to Einstein equations
have been proposed. In particular, the distinction between
``hard'' and ``soft'' singularities has been repeatedly
emphasized, and many contributions of a worthy value from the
\emph{geometric} standpoint have been given. We think, however,
that Hilbert's simple definition of the regularity of a metric
tensor is chiefly preferable for \emph{physical} reasons.

\vskip0.80cm
\par A friendly epistolary discussion with Prof. G.
Morpurgo is gratefully acknowledged.


\vskip0.80cm \small
\begin{thebibliography}{9}

\bibitem{1}
A. Loinger and T. Marsico, \emph{arXiv:0706.3891 v3}
$[$physics.gen-ph$]$ 16 Jul 2007.

\bibitem{2}
D. Hilbert, \emph{Mathem. Annalen}, \textbf{92} (1924) 1; also in
\emph{Gesammelte Abhandlungen}, Dritter Band (J. Springer, Berlin)
1935, p.258. This memoir reproduces previous academic
communications: \emph{G\"ott Nachr.} -- Erste Mitteilung,
vorgelegt am 20. Nov. 1915; zweite Mitteilung, vorgelegt am 23.
Dez. 1916; \emph{G\"ott Nachr.}, vorgelegt am 25. Jan. 1918.

\bibitem{3}
See: K. Schwarzschild, \emph{Berl. Ber.}, (1916) 424; S. Weinberg,
\emph{Gravitation and Cosmology etc.} (Wiley and Sons, New York,
\emph{etc}.) 1972, Chapt.\textbf{11}, sect. \textbf{6}.

\bibitem{4}
A. Einstein, \emph{Ann. Math.}, \textbf{40} (1939) 922.

\bibitem{5}
Cf., \emph{e.g.}, Y. Wang and P. Mukherjee,
\emph{arXiv:astro-ph/0703780 v1} -- 21 March 2007, and references
therein.

\bibitem{6}
A. Loinger, \emph{arXiv:physics/0504018 v1} -- 3 April 2005, and
references therein.

\end{thebibliography}
\end{document}